# Upgrading Urban Water Storage System: Achieving Water Conservation, Power Generation, Carbon Reduction, and Water Quality Enhancement

The Affiliated High School of Peking University
Zhang Xiao
07/08/2023

## Abstract：

　　Beijing is a megacity with a population of over 20 million. It has a massive demand for water and electricity and is responsible for achieving carbon neutrality. However, Beijing is in a highly arid region, and its local electricity production supply is less than 30% of the demand. The city's carbon neutrality goal also faces enormous challenges. To address these problems, this paper proposes a plan to cover the Beijing area's surface reservoirs and water channels with solar panels. The results show that the plan is expected to save an average of 383 million $m^3$ of water per year (which can meet the annual water needs of 2.13 million households), generate 72.6 billion kWh of electricity (which can meet the annual electricity needs of 25.23 million households), reduce carbon dioxide emissions by 6.84 billion kg (equivalent to the annual carbon dioxide absorption of 3.1 billion mature trees), and reduce the concentration of one carcinogenic substance, microcystin, by 97.3%, while also addressing the issue of bromate. The results strongly suggest the potential social and economic value of using solar panels to cover surface reservoirs and water channels. Field control experiments and related research should be actively promoted.

## Introduction：

　　According to the World Bank, the global per capita freshwater resource in 2020 was about 5,050 cubic meters per person[1]. However, Beijing's per capita freshwater resource is only 118 cubic meters per person[2], which is extremely scarce. Beijing has taken several measures to address the water shortage, including the South-to-North Water Diversion Project and public appeals for water conservation. This paper explores another approach to water conservation: reducing water loss during urban water storage and transportation. A study by the Chinese

Academy of Sciences in 2020 found that the lakes in the Tibetan Plateau evaporate an amount of water equivalent to 3,570 West Lakes per year[3]. This data suggests that water evaporation is a serious way of water loss during water storage. Another study conducted abroad found that covering the surface can reduce water evaporation by 81-83%[4]. These two studies provide data support for the water conservation method proposed in this paper: installing opaque covers on the surface of Beijing's reservoirs and canals to achieve water conservation.

According to data released by the Beijing Municipal Government, Beijing's electricity consumption reached 114 billion kWh in 2020[5], while local electricity production met only 30% of the demand[6]. With the development of the economy and the popularization of electric vehicles, electricity demand is expected to continue to rise. According to data released by the National Energy Administration, Beijing's average annual solar radiation is 1400-1750 kWh/m2, which is in the "very rich" belt of solar radiation resources[6]. This paper proposes to install opaque covers on surface reservoirs and canals; when solar panels are used as covers, the idle space on the water surface can be fully utilized for solar power generation to increase Beijing's electricity supply.

According to a report by the United Nations, the world needs to reduce greenhouse gas emissions to address global climate change[8]. Data from the National Bureau of Statistics show that in 2019, 97.52% of local electricity generation in Beijing still came from fired power plants[9]. The baseline emission factor for the North China power grid (including Beijing) is 0.9419 kg CO2/kWh[10], which shows that fired power plants in Beijing still emit a large amount of carbon dioxide to society. The 2022 Beijing Energy Development Plan for the 14th Five-Year Plan proposes to "implement renewable energy substitution actions and accelerate the development and utilization of local renewable electricity"[11]. This paper proposes to install opaque solar panels on surface reservoirs and canals, using solar power generation to replace traditional fired power generation. This is consistent with the direction of the Beijing Municipal Government's plan and can effectively reduce carbon dioxide emissions.

According to a study by the Chinese Academy of Sciences, China is one of the countries with the broadest distribution and most profound impact of harmful algae, and high concentrations of cyanobacteria cells and the microcystins they produce can threaten drinking water's safety[12]. Microcystins can harm the liver, kidneys, gonads, and nervous system and are incredibly toxic to mammals. They often cause irreversible pathological changes in minutes or hours[13]. A study published in the journal Environmental Science & Health found that the concentration of microcystin toxins in the Guanting Reservoir in Beijing reached a maximum of 1.231 (μg/L) in 2014, exceeding the national standard by 23.1%[14]. This study result suggests that Beijing has microcystins affecting water safety and poses a threat to public health. Microcystins are intracellular toxins that are synthesized in cyanobacteria cells. They are released into the water only when the cells die and rupture[12]. This paper proposes to install opaque solar panels on surface reservoirs and canals. By blocking sunlight, solar panels can inhibit the photosynthesis of cyanobacteria and other harmful algae, thereby preventing their growth and reproduction and significantly reducing the concentration of microcystins in the water.

According to a study by the Chinese Academy of Sciences Ecological Environment Research Center, water samples were collected from three water plants in Beijing every two months, and the bromate in the three water plants was found to be mainly from surface water contamination[15]. The study results show that there is indeed a problem of bromate contamination in surface water, such as reservoirs in Beijing. Bromide is a harmless compound naturally present in all water bodies. However, when water is exposed to sunlight, bromide reacts with chlorine to produce toxic and carcinogenic bromate[16]. This paper proposes to install opaque solar panels on surface reservoirs and canals, which can significantly reduce the production of bromate by blocking sunlight.

In conclusion, this paper proposes to install opaque solar panels on surface reservoirs and canals, which can effectively achieve the goals of water conservation, power generation, carbon reduction, and water quality improvement.

**Program design and calculations:**

The recognized water evaporation model is the Dalton evaporation model ( $E = \alpha \frac{(e_s - e_q)}{P}$ ). There are several formulas for calculating the evaporation coefficient α, including the original Soviet power plant design institute formula, the Shi Chengxi formula, the R-H formula, and the Adams formula. The universal formula used in China is[17]:

$$\alpha = [2.77 + 1.56w^2 + 0.25(T_s - T_q)]^{\frac{1}{2}} \times 10^{-1} \quad mm \cdot d^{-1}(hPa)^{-1}$$

We can see that the factors affecting water evaporation include wind speed, water temperature, air temperature, relative humidity, saturated vapor pressure difference, and atmospheric pressure.

The design proposed in this paper is to lay solar panels (with an insulation layer underneath) on the surface of the reservoir and canal 0.5 meters above the designed highest water level. This will cover the water surface to block sunlight, slow down the wind speed on the surface, and reduce the temperature difference between water and air and the saturated vapor pressure difference. In addition, when laying the solar panels, it is necessary to reserve the entrance and exit for rainwater and overflow and the channel for engineering construction and maintenance. (Note: Because Beijing is in the northern hemisphere at a latitude of about 39.90°, solar panels need to be tilted at a certain angle to obtain the best power generation efficiency. The flat-laying solar panel scheme proposed in this paper will reduce the power generation efficiency by about 16%[23]. However, it can minimize water evaporation and maximize the reduction of harmful algae and bromate content.)

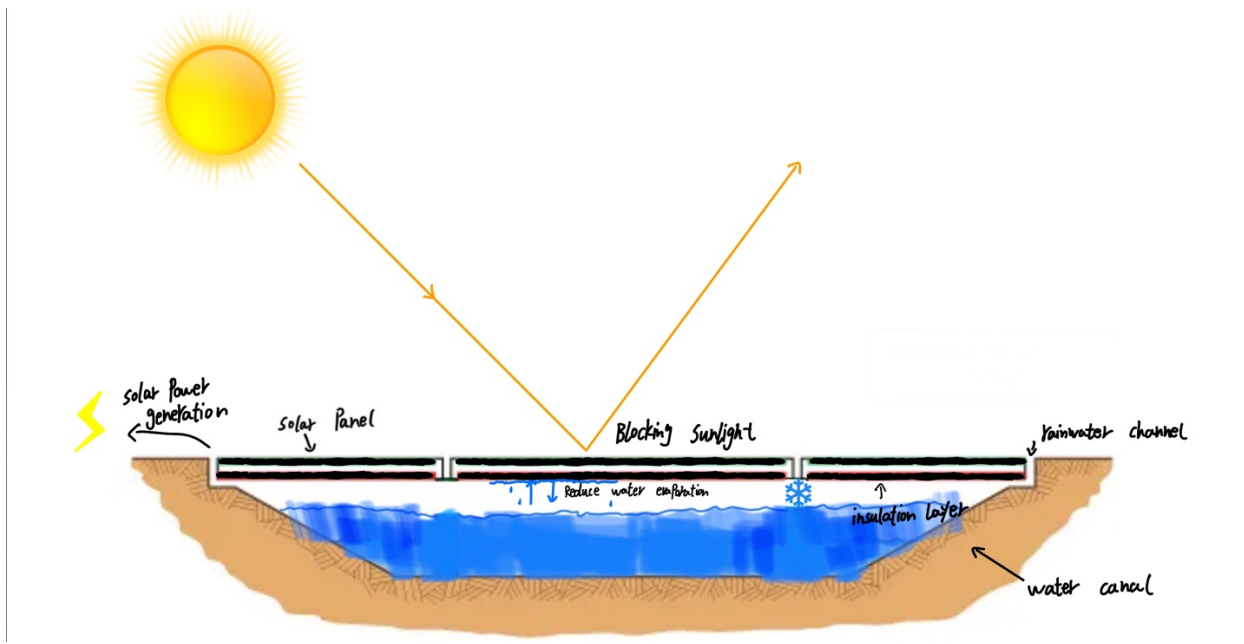

(Fig.1 Schematic diagram of the program)

British physicist and chemist John Dalton proposed that water's saturated water vapor pressure is greater than the water vapor pressure in the air, which promotes water vapor diffusion from the water's surface to the air. This causes the water vapor pressure in the thin layer of air near the water surface to decrease. In order to maintain water vapor saturation on the water surface, new water molecules evaporate from the water body and are replenished in the air on the water surface. This is the cause of water evaporation. The increase in wind speed helps to diffuse and transport water vapor. In contrast, the increase in atmospheric pressure increases the air density at the water's surface, making it difficult for water molecules to escape from the water's surface. Therefore, the amount of water evaporation should be proportional to the saturated water vapor pressure and increase with the increase in wind speed but decrease with the increase in atmospheric pressure. The famous Dalton evaporation law is a concise statement of the cause and quantitative principle of water evaporation on the surface[18].

So, the author set up an experimental environment, including covering the canal model with solar panels, storing electricity with a lithium battery pack, and designing and manufacturing hardware and circuits that collect temperature, humidity, wind speed, air pressure, ultrasonic sensors, and power management. The author also wrote a Python program to collect real-time data (including temperature, humidity, wind speed, air pressure, and water level changes), display data in real-time, display historical data trends, calculate water evaporation, monitor, record water level, and other functions. The integrated hardware and software design allows for direct and accurate observation of the effects of temperature, humidity, wind speed, and air pressure on water evaporation. The experimental environment can also study and test data on water evaporation, solar power generation, microcystins, and bromate content.

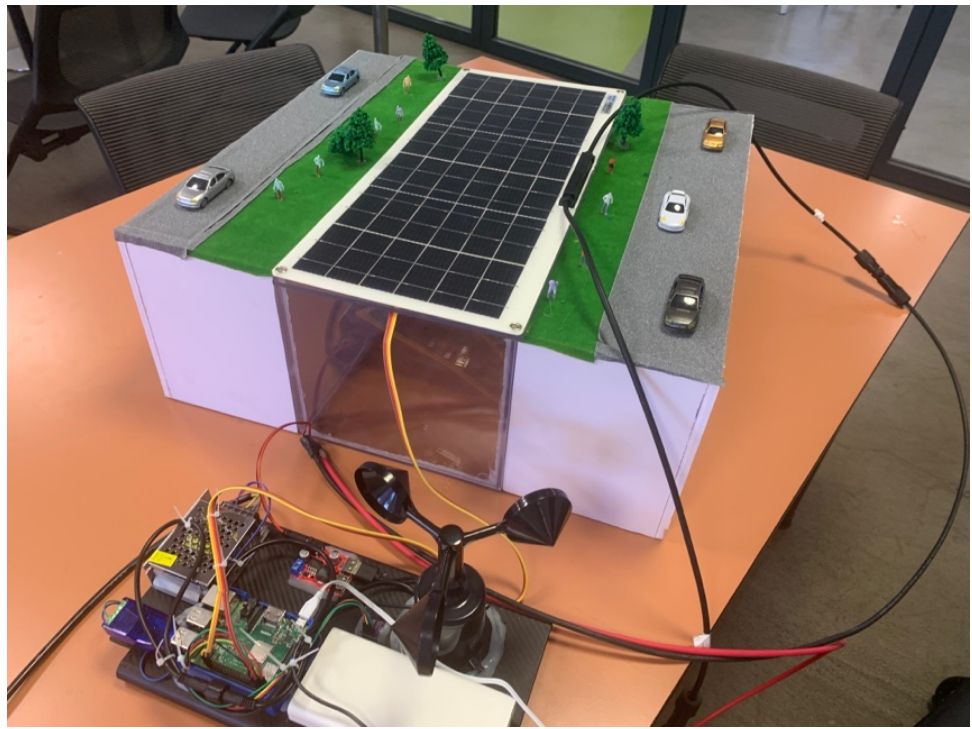

(Fig.2 Experimental Models)

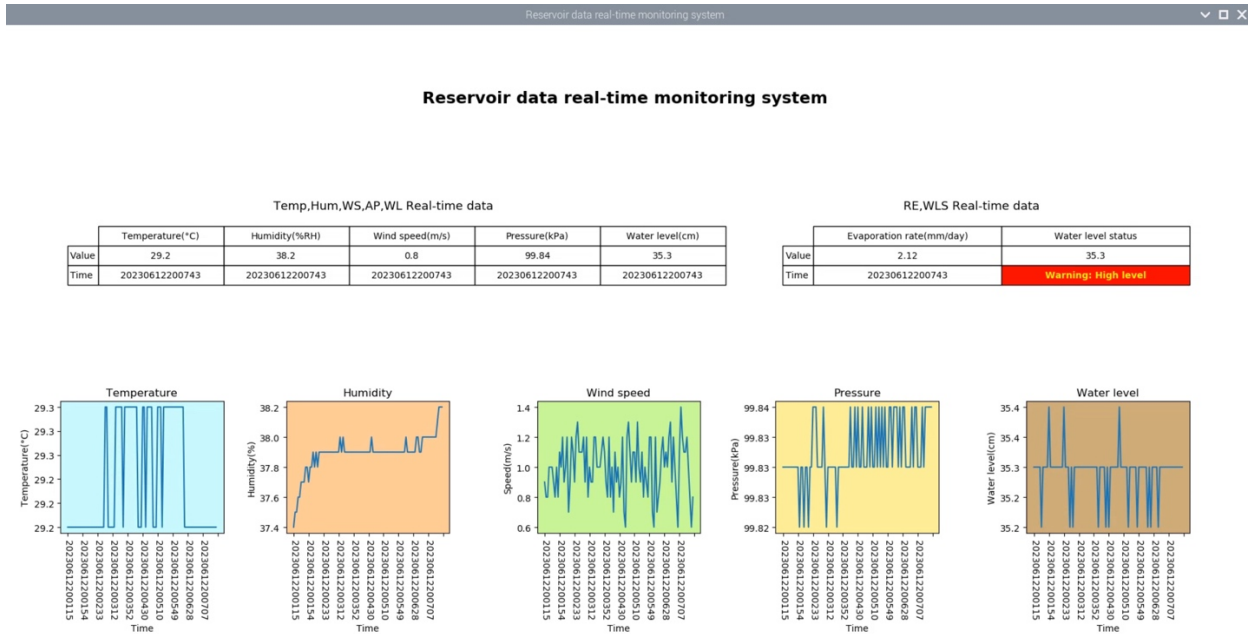

(Fig.3 Real-time data graphs)

The data on water conservation, power generation, carbon reduction, and water quality improvement are first calculated using data released by the government and research institutions and then calculated using the measured data in the experimental environment that the authors set up.

Regarding water conservation, the data from the Chinese Academy of Sciences Institute of Geographical Sciences and Resources shows that the average annual water surface evaporation in North China (including Beijing) is 1699.5mm[19]; the data released by the Beijing Water Authority shows that the surface area of the Beijing surface reservoir is 274.56km2[20]; (note: because Beijing Water Authority was unable to provide relevant data on the canal area in the government information disclosure application, this paper is currently only using the surface reservoir area for calculation), according to these two data, it can be calculated that the average annual reservoir water evaporation in Beijing is about 467 million m³, $1.6995m \times 2.7456 \times 10^8 = 4.67 \times 10^8 \ m^3$. According to a study based on covering the reservoir with a tarp to prevent water evaporation, covering the water surface with a tarp can reduce water evaporation by 81%-83%[4] (median of 82%), since the location of this study is close to the latitude of Beijing and the climate environment is similar, so if the Beijing reservoir is paved with solar panels, referring to the data of the study results, it can be calculated that the Beijing reservoir can expect to reduce about 383 million m³ of water evaporation per year, $4.67 \times 10^8\ m^3 \times 82\% = 3.8294 \times 10^8\ m^3$. The saved water can be used by 2.13 million Beijing households annually. (Calculated according to the first-tier annual water use of Beijing households of 180 m³/household[21].)

Regarding power generation, according to the data of the National Energy Administration, the average annual solar radiation in Beijing is 1400-1750 kWh/m²[7] (median of 1575), and the current mainstream solar panel conversion efficiency is 18%-22%[22] (median of 20%). According to the China Meteorological Administration Solar and Wind Energy Center data, when solar panels are laid flat in Beijing, the average solar radiation they receive is about 84% of the solar radiation they receive at the optimal tilt angle[23]. It is known that the surface area of Beijing's surface reservoirs is 274.56[19] km²[20]. According to the above data, suppose solar panels are laid flat on the surface of Beijing's reservoirs. In that case, it can be calculated that Beijing's annual average solar power generation can be about 7.2649×10¹⁰ kWh, $1575 \times 20\% \times 2.7456 \times 10^8 \times 84\% = 7.2649 \times 10^{10}\ kWh$, which 25.23 million households in Beijing can use throughout the year. (Calculated according to the first-tier annual power consumption of Beijing households of 2880kWh/household[24].)

Regarding carbon reduction, according to the Ministry of Ecology and Environment data, the baseline emission factor for the North China regional power grid (including Beijing) is 0.9419 kg of carbon dioxide per kWh[9]. If we calculate based on the above solar power generation of 7.2649×10¹⁰ kWh, Beijing can reduce about 6.8428×10¹⁰ kg of carbon dioxide emissions per year. The reduced carbon dioxide emissions are equivalent to the carbon dioxide absorbed by 3.1 billion mature trees yearly. (According to the data of the European Environment Agency: a mature tree can absorb 22 kg of carbon dioxide per year[25].)

Regarding water quality improvement, since cyanobacteria need photosynthesis to grow and reproduce, if the light is blocked, it can theoretically prevent the growth of cyanobacteria. Microcystin is synthesized in the cells of cyanobacteria, so if there are no cyanobacteria, there will naturally be no microcystin. Therefore, covering the reservoir's water surface with solar panels can create an environment where cyanobacteria cannot grow and reproduce, thereby eliminating or significantly reducing the production of microcystin. In addition, naturally occurring bromide and chlorine in water react to produce bromate when exposed to sunlight. The reaction process is as follows:

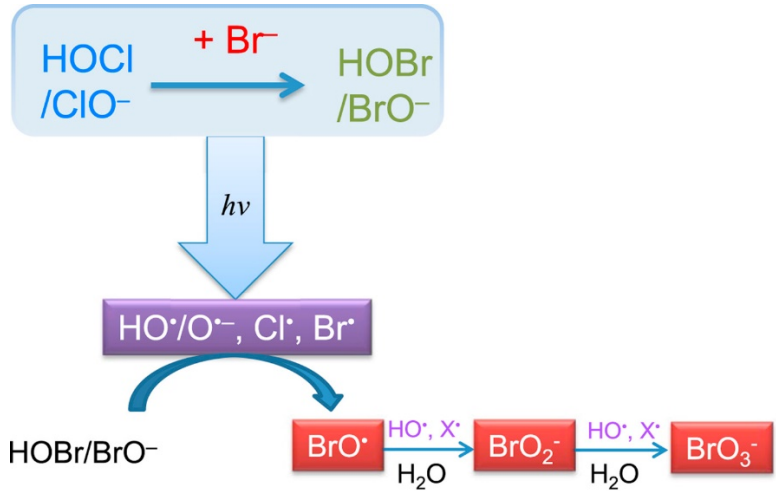

(From https://www.sciencedirect.com/science/article/abs/pii/S0045653517308421)

So, covering the reservoir's water surface entirely with solar panels could stop or dramatically reduce bromate production by blocking sunlight.

    The following arithmetic is based on the experimental environment built by myself and the data obtained from the experiment:

    **Experimental design**: Two identical water containers A and B were used. The outer periphery of the water container was opaque. The experimental group A and the control group B were respectively injected with 10L of laboratory pure water, 100ml of cyanobacteria and green algae mixture (solid content 5%), 1g of nitrogen phosphorus potassium nutrients (N-P2O5-K2O, ratio 15-5-25), and a piece of 20*40cm ecological cotton for algae attachment. The experimental group A and the control group B were placed outdoors and exposed to the sun completely. A non-transparent plate was covered in experimental group A, with about 0.5cm from the top of the container, so that the air could still flow. The top of the control group B was completely open and exposed to the sun. The experiment started on June 21, 2023, and ended on June 30, 2023, for ten days outdoors. According to the data released by the Beijing Meteorological Observatory, the average temperature during this period was 31.1-21.7°C, the average relative humidity was 40.4%, and the wind speed was all <3 Degrees.

    The results of the experiment are as follows:

    **Water level change：**

On the first day, groups A and B's initial water levels were 150mm. On the tenth day, the water level of group A was 102mm, and the average daily evaporation rate was 4.8mm/day. The water level of group B was 33mm, and the average daily evaporation rate was 11.7mm/day. The results showed that group A reduced the water evaporation rate by 58.97% by adding a non-transparent cover.

**Microcystins content change：**

The school did not have the instruments and reagents to detect microcystins' content accurately. However, microcystins are produced by cyanobacteria. Therefore, the author speculates on the change in microcystin content by observing cyanobacteria's growth and reproduction. On the first day, 10 10mm ecological cotton was cut from groups A and B and mixed with 100 ml water to obtain algae samples. The initial number of cyanobacteria in the block of group A water sample was 16, and the initial number of cyanobacteria in the block of group B water sample was 15. On the fourth day, 10*10mm ecological cotton was cut from groups A and B and mixed with 100 ml water to obtain algae samples. The initial number of cyanobacteria in the block of group A water sample was averaged to be 17, and the initial number of cyanobacteria in the block of group B water sample was averaged to be 52. The results showed that group A reduced the growth and reproduction of cyanobacteria by 97.3% compared with group B, and the theoretical reduction of microcystins content was 97.3%.

**Bromide content change：**

The school did not have the instruments to detect the bromide content accurately. However, the bromide content can be roughly estimated by the color change of the solution after the reagent and the water sample are mixed. Therefore, the author used this method to verify the influence of sunlight on bromide production. The preparation method of the reagent and the detection method are as follows:

Prepare LR1 solution and LR2 solution. LR1 solution is a hydrochloric acid solution (concentrated hydrochloric acid mixed with distilled water in a volume ratio 1:4), and LR2 solution is a 1% potassium iodide solution.

Mix equal LR1 and LR2 solution volumes into a test tube and add the water sample.

After 5 minutes, observe the color change of the solution. If bromide is detected, the solution will turn brown-yellow; the higher the content, the darker the color.

On the first day, water samples were taken from groups A and B and mixed with the reagent (each test tube contains: 3ml LR1 solution, 3mlL LR2 solution, and 1ml water sample). The mixed solutions of groups A and B were both colorless. On the fourth day, water samples were taken from groups A and B and mixed with the reagent (each test tube contains: 3ml LR1 solution, 3mlL LR2 solution, and 1ml water sample). The mixed solutions of groups A and B were still colorless. On the seventh day, water samples were taken from groups A and B and mixed with the reagent (each test tube contains: 3ml LR1 solution, 3mlL LR2 solution, and 1ml water sample). The mixed solution of group A was still colorless, but the mixed solution of group B turned brown-yellow.

The results showed that group A effectively inhibited bromide production by adding a non-transparent cover. However, the exact reduction ratio could not be calculated due to the need for accurate bromide content data.

**Solar panel power generation:**

The solar panel installed on the water channel model is a monocrystalline flexible solar panel with a nominal conversion efficiency of 20%, peak power of 20W, and output voltage of 18V. According to the actual test on a sunny day, the solar panel facing south at an angle of 45 degrees can charge a 20000mAh (5V) power bank for about 60% of the day. When the installation angle is 0 degrees (flat), the solar panel can charge a 20000mAh (5V) power bank for about 50% of the day. The results show that the charging efficiency of the flat solar panel will be reduced by about 16.77%.

(Note: Due to the overlap of the 10-day experiment on water evaporation, microcystin, and bromide with school exams and other necessary activities, it was impossible to conduct continuous experiments and record the results. Therefore, the algae count was only conducted on the first and fourth days of the experiment; bromide was not observed to change color on the first and fourth days, and the experiment was repeated on the seventh day; the water level was recorded on the first day and the last day, and the maximum value of the experiment was recorded after ten days. So, we can see that experiment is not sufficient due to the limitations of time, instruments, and reagents, but the results can still verify the theoretical inference.)

# Discussion:

Both the use of government and research institution data and our experimental data demonstrate that covering solar panels on water surfaces can achieve the goals of water conservation, power generation, carbon reduction, and water quality improvement.

In the experiments, the evaporation rate in experimental group A decreased by 58.97%, much lower than the 81%-83% data reported by research institutions. This may be due to the small volume of the container and its opaque, dark-colored outer wall. Under sunlight, the water temperature inside the container can significantly exceed the actual temperature in the reservoir, leading to higher evaporation in experimental group A (gaps between the cover and the container allow water vapor to escape). Moreover, due to the limitations of the experimental instruments, accurate and specific data on algal toxins and bromate concentrations could not be obtained. Based on the above conclusions, it is recommended to actively promote field research by digging multiple water pools measuring 100 m$^2$ and 3 m deep near the Guanting Reservoir or Miyun Reservoir. Solar panels should be installed on some pools (laid flat or at different angles) for comparative experiments in real-field conditions to obtain accurate data on evaporation, power generation, carbon reduction, microcystins, and bromate concentrations. Urban water storage systems should be upgraded as soon as possible to benefit society and the Earth.

**Acknowledgments:**